\documentclass[twocolumn,showpacs,amsmath,amssymb]{revtex4}
\usepackage{amsmath} %

\usepackage{graphicx}
\usepackage{amsfonts}
\usepackage{dcolumn}
\usepackage{bm}

\newcommand \dCG{$d$-CG\ }
\newcommand \sCG{$s$-CG\ }

\begin{document}

\title{Coarse-grained Monte
Carlo simulations of the phase transition of Potts model on weighted
networks}
\author{Chuansheng Shen$^{1,2}$}

\author{Hanshuang Chen$^{1}$}

\author{Zhonghuai Hou$^{1}$}\email{hzhlj@ustc.edu.cn}

\author{Houwen Xin$^{1}$}

\affiliation{$^{1}$ Hefei National Laboratory for Physical Sciences at Microscales \& Department of Chemical Physics, University of
 Science and Technology of China, Hefei, 230026, China \\
 $^{2}$Department of Physics, Anqing Teachers College, Anqing, 246011, China}


\begin{abstract}
Developing effective coarse grained (CG) approach is a promising way
for studying dynamics on large size networks. In the present work,
we have proposed a strength-based CG (\sCG) method to study critical
phenomena of the Potts model on weighted complex networks. By
merging nodes with close strength together, the original network is
reduced to a CG-network with much smaller size, on which the
CG-Hamiltonian can be well-defined. In particular, we make error
analysis and show that our strength-based CG approach satisfies the
condition of statistical consistency, which demands that the
equilibrium probability distribution of the CG-model matches that of
the microscopic counterpart. Extensive numerical simulations are
performed on scale-free networks, without or with
strength-correlation, showing that this \sCG approach works very
well in reproducing the phase diagrams, fluctuations, and finite
size effects of the microscopic model, while the \dCG approach
proposed in our recent work [Phys. Rev. E 82, 011107(2010)] does
not.

\end{abstract}
 \maketitle

\section{Introduction}
In the last two decades, we have witnessed dramatic advances in
complex networks research, which has been one of the most active
topics in statistical physics and closely related disciplines
\cite{RMP02000047,AIP02001079,SIR03000167,PRP06000175,PRP08000093}.
The central issue in this field is to study how the topology of
networks influences dynamics, such as phase transition,
self-organized criticality and epidemic spreading, etc. Usually,
Monte Carlo (MC) simulations \cite{Lan2000} have been widely used to
study such dynamics. However, the sizes of many real-world networks
are very large, such as human brain composed of about $10^{11}$
neurons and $10^{14}$ synapses \cite{RMP06001213}, and thereby
brute-force simulations are quite expensive and sometimes even
become impossible. Phenomenological models, such as mean-field
description, may capture certain properties of the system, but often
ignore microscopic details and fluctuation effects which may be
important near some critical points. Therefore, a promising way to
bridge the gap between the microscopic details and system level
behaviors is to develop coarse-grained (CG) approaches, aiming at
significantly reducing the degree of freedom while properly
preserving the microscopic information of interest.

Recently, several CG approaches have been proposed in the
literature. Renormalization transformation has been used to reduce
the size of self-similar networks while preserving the most relevant
topological properties of the original ones
\cite{PRL04016701,NTR05000392,PRL06018701,PRL08148701}. Gfeller and
Rios proposed spectral decomposition technique to obtain a
CG-network which can reproduce the random walk and synchronization
dynamics of the original network \cite{PRL08174104}. Kevrekidis et
al. developed equation-free multiscale computational methods to
accelerate simulation using a coarse time-stepper
\cite{CMS03000715}, which has been successfully applied to study the
CG dynamics of oscillator networks \cite{PRL06144101}, gene
regulatory networks \cite{JCP06084106}, and adaptive epidemic
networks \cite{EPL08038004}. Nevertheless, none of the works
mentioned above has considered critical phenomena in complex
networks, which has been a frontier topic in the context of network
science \cite{RMP08001275}.

Very recently, we have proposed a degree-based CG (\dCG) approach to
study the critical phenomena of the Ising model and the SIS-epidemic model in unweighted networks \cite{PRE10011107}. A local mean field (LMF) scheme was introduced to generate the CG network from the microscopic one. Specifically, we have proposed a so-called condition of statistical consistency (CSC) that the CG-model should satisfy to guarantee the validity of the
CG-approach. We showed that the CSC can be exactly fulfilled if we merge nodes with the same degree together. Extensive
numerical simulations showed that our \dCG approach does work very well to reproduce the phase transition behaviors of the
original network, including the critical point and the fluctuation properties, but with much less computational efforts. Our
method also makes it feasible to investigate the finite size effects of both models, which should be much more expensive and
even forbidden if we use brute-force methods. However, this \dCG approach can only apply to binary networks, i.e., each of the
link in the network either exists or not, but with no weight. As we know, many real-world networks are intrinsically weighted,
with their links having diverse strengths. Examples include the collaboration networks \cite{PRE01016131,PRE01016132,PHA02000590}, airport networks
\cite{PNA04003747,PRE05036124}, metabolic networks
\cite{NAT04000839}  predator-prey relationship networks
\cite{Pin2002}, and so on. Therefore, a straightforward question is: Can we use CG approaches to study the critical phenomena in weighted
networks?

   To answer this question, in the present work, we have considered the critical phenomena of the Potts model in weighted
complex networks. The Potts model is related to a number of
important topics in statistical and mathematical physics
\cite{RMP82000235,Gar1979} and was successfully applied to neural
networks, multiclass classification problems, graph coloring
problem, and so on. It contains a system of coupled nodes, each of
which has \emph{p} possible states. Only when two nodes are in the
same state, they have pairwise interactions. With the increment of
temperature, the Potts model undergoes an order-disorder phase
transition at some critical temperature. For $p = 2$, Potts model is
equivalent to the well-known Ising model. Instead of the \dCG
scheme, we have proposed a strength-based CG (\sCG) approach, where
those nodes with \emph{similar} strength are merged together to form
a CG-node. Note that in weighted networks, it is unpractical to
merge nodes with exactly the same strength together. By detailed
analysis of the discrepancy between the Hamiltonian of a CG
configuration and that of its corresponding microscopic
configurations, we show that the \sCG approach can approximately
satisfy the CSC defined on weighted networks. Extensive numerical
simulations are performed on scale-free(SF) networks, without or
with strength-correlation, showing that our \sCG approach works very
well in reproducing the phase diagrams, fluctuations, and finite
size effects of the microscopic model, while the simple \dCG does
not. Compared to our previous work  \cite{PRE10011107}, the present
study step forward several important steps. First of all, we should
note that \sCG is a brand new method compared to \dCG and the latter
cannot apply to weighted networks, although they share some similar
ideas. Secondly, weighted networks are of more ubiquitous importance
than binary unweighted ones, thus the \sCG approach should find more
applications.  What is more, we have extended the study from the
simple two-state Ising model to a more general one, the multi-state
Potts model. In addition, we have performed error analysis in the
present study, which clearly demonstrates the robustness of our
approaches.

\section{Coarse Graining Procedure} \label{sec2}

\subsection{CG Potts Model}
In this paper, we consider the $p$-states Potts model on a weighted
network consisted of $N$ nodes, whose Hamiltonian is given by
\begin{equation}
H =  - \sum\limits_{i<j} {w_{ij} \delta _{\alpha _i ,\alpha _j } } ,
\label{eq1}
\end{equation}
where $w_{ij}$ is the weight on the edge connecting a pair of nodes
$i$ and $j$ ($w_{ij}=0$ if the nodes $i$ and $j$ are not connected).
$\alpha _i(= 1,\cdots,p)$ denotes the state of node $i$, $\delta
_{\alpha_i,\alpha_j } = 1$ if $\alpha_i=\alpha_j$ and $0$ otherwise.

To setup the CG-Potts model, one needs to obtain the CG-Hamiltonian
defined on the CG-network, followed by CG-MC simulations to study
the dynamic behaviors. The CG-network is simply obtained by
node-merging, i.e., $q_\mu$ nodes within the original micro-network
are merged into a single CG-node $C_\mu$, where $\mu=1,...,N^c$
labels the CG-node and $N^c$ is size of the CG-network. Following
the LMF scheme used in Ref. \cite{PRE10011107}, the weight of link between two CG nodes $\mu$ and $\nu$ reads,
\begin{equation}\label{eqCGA}
 \bar w_{\mu\nu} =
  \begin{cases}
   \frac{2}{q_\mu(q_\mu-1)}\sum\limits_{i,j \in C_\mu; i<j} w_{ij}
  & \text{for $\mu = \nu$},  \\
  \frac{1}{q_\mu q_\nu}\sum\limits_{i \in C_\mu,j \in C_\nu} w_{ij}
  & \text{for $\mu \ne \nu$}.
 \end{cases}
\end{equation}

The CG-Hamiltonian $\bar H$ can be readily obtained,
$$ \bar H = \bar H_1 + \bar H_2 $$
where
\begin{subequations}\label{eqCGH}
\begin{eqnarray}
 \bar H_1 &=& - \sum\limits_\mu {\bar w_{\mu\mu} } \sum\limits_\alpha  {\frac{\eta _{\mu,\alpha} (\eta_{\mu,\alpha}  - 1)}{2}}
\label{eq4a} \\
 \bar H_2 &=& - \sum\limits_{\mu,\nu( > \mu)} {\bar w_{\mu\nu} } \sum\limits_\alpha  {\eta _{\mu,\alpha} \eta _{\nu,\alpha} }
\label{eq4b}
 \end{eqnarray}
\end{subequations}
Herein, $\bar H_1$($\bar H_2$) denote CG interactions inside(among)
the CG-nodes, respectively. $\eta_{\mu,\alpha}$ stands for the
number of $\alpha$-state micro-nodes inside $C_\mu$. Since there are
$\frac{{\eta _{\mu,\alpha} (\eta _{\mu,\alpha} - 1)}}{2}$ possible
distinct pairs of $\alpha$-state micro-nodes inside $C_\mu$, and
each pair has a weighted coupling $w_{\mu\mu}$, the CG-interactions
among all the $\alpha$-state nodes inside $C_\mu$ is given by
$$ \bar H_{\mu,1}^{(\alpha)} = - \bar w_{\mu\mu} \frac{{\eta _{\mu,\alpha} (\eta _{\mu,\alpha} - 1)}}{2}. $$ Summation this over
all CG-nodes
$\mu$ and states $\alpha$ gives the result in Eq.(3a). Eq.(3b) can
be interpreted in a similar way. Note that Eq.(3) are closed at the CG level, i.e., as long as one has constructed the CG-network, $\bar {\bm w}$ and $\bar H$ are then
both well defined, based on which one can perform CG-MC simulations
without going back to the micro-level.

\subsection{CSC: Condition of statistical consistency}

The above procedure tells us how to calculate the CG-Hamiltonian
if we already have the CG-network. However, which $q_\mu$ nodes are
merged together to form a CG-node $C_\mu$ is yet not determined.
Generally speaking, one may construct the CG-network deliberately,
for instance, one may simply generate $N^c$ values, $q_\mu$ obeying
$\sum_{\mu=1}^{N^c} q_\mu=N$ and then just randomly merge $q_\mu$
micro-nodes to form $C_\mu$. Therefore, an important question
arises: How to guarantee that the CG-model can reproduce the
dynamics of the corresponding microscopic model correctly?

We address this problem by extending the so-called CSC as proposed
in \cite{PRE10011107}. We introduce $\vec
\eta_\mu=\{\eta_{\mu,\alpha}\}_{\alpha=1,..., p}$ to denote the
state of $C_\mu$ and $\vec {\bm \eta}= \{\vec
\eta_\mu\}_{\mu=1,...,N^c}$ to denote the configuration of the
CG-network. Note that a given CG configuration $\vec {\bm \eta}$
corresponds to many microscopic configurations, which defines the
degeneracy factor $g(\vec {\bm \eta})$. In the equilibrium state of
the CG-model, the probability of finding a given CG-configuration
$\vec {\bm \eta}$ is given by the canonical distribution, i.e.,
$$p_{\rm CG}(\vec {\bm \eta})=g(\vec {\bm \eta})e^{-\bar H/k_BT}/\bar Z, $$
where $\bar Z=\sum_{\vec {\bm \eta}} p_{\rm CG}(\vec {\bm \eta}) $
is the CG partition function. It is important to note, however, that
$p(\vec {\bm \eta})$ can be calculated exactly from the equilibrium
distribution of the micro-model,
$$ p_{\rm micro}(\vec {\bm \eta}) = \sum\nolimits_{}' e^{ - H/k_B T}/Z,$$
where $Z$ is the partition function of the micro-model, and the
prime means summation over all the microscopic configurations that
contribute to $\vec {\bm \eta}$. Since we are interested in the
equilibrium phase transition behavior of the Potts model, we thus
assert that for the CG-model to be statistically consistent with the
micro-model, $p_{\rm CG}(\vec {\bm \eta})$ and $ p_{\rm micro}(\vec
{\bm \eta})$ must be equal, i.e., the CSC reads
\begin{equation} \label{eqCSC}
g(\vec {\bm \eta})e^{-\bar H/k_BT}/\bar Z =\sum\nolimits_{}' e^{ -
H/k_B T}/Z.
\end{equation}

\subsection{\sCG Scheme and error analysis}

 In the present work, we propose a \sCG
scheme to construct the CG-network, i.e., nodes with same or similar
strengths are merged together to form a CG-node, where the strength
$s_i$ of node $i$ is defined as $s_i  = \sum\nolimits_j {w_{ij} } $
\cite{PRL01005835,PNA04003747}. In the following, we will show that
if nodes inside each CG-node have same strengths, the CSC will hold
exactly within the ANA. In addition, if the strengths within $C_\mu$
are nearly the same, the CSC can also hold approximately.

In the literature, ANA
\cite{RMP08001275,PRE03036112,PRL02258702,PRE02035108} has been
widely used to study the ensemble averaged dynamics of
 complex networks and proved to be
successful. ANA assumes that one can replace the dynamics on a given
network by that on a weighted fully connected graph with
connectivity $ A_{ij}=d_i d_j/(D N)$, where $d_i$ ($d_j$) denotes
the degree of node $i$ ($j$) and $D$ is the mean degree of the
network. Analogously, in weighted networks link weight can be
expressed as
\begin{equation}
 w_{ij}=s_i
s_j/(S N)
   \label{eqANA}
\end{equation}
where $S$ is the mean strength of the network. Substituting
Eq.(\ref{eqANA}) into Eq.(\ref{eqCGA}), the adjacency matrix of the
CG-network now reads,

\begin{subequations}\label{eqCGA_ANA}
\begin{eqnarray}
\bar w_{\mu\mu} &=& {\frac{2}{{q_\mu (q_\mu  - 1)}}} \sum\limits_{i
< j \in C_\mu } {\frac{{(S_\mu  + \delta s_i )(S_\mu  + \delta s_j
)}}{{S N}}} \nonumber \\
&=& \frac{S_\mu^2}{{S N}} \left( {1 - \Omega_\mu } \right)
\\
 \bar w_{\mu\nu} &=& {\frac{1}{{q_\mu q_\nu}}}\sum\limits_{i \in C_\mu ,j \in C_\nu } {\frac{{(S_\mu  + \delta s_i )(S_\nu  +
\delta s_j )}}{{S N}}} \nonumber \\
& =& \frac{1}{{S N}}\sum\limits_{i \in C_\mu ,j \in C_\nu } {S_\mu
S_\nu }
  = \frac{S_\mu S_\nu}{{S N}}
 \end{eqnarray}
\end{subequations}

Herein, we have written $s_i=S_\mu+\delta s_i$, with
$S_\mu=\frac{1}{q_\mu}\sum_{i\in C_\mu}s_i$ being the mean strength
within $C_\mu$. $\Omega_\mu = \frac{{ \left< \delta s^2 \right
>_\mu }}{S_\mu^2 (q_\mu  - 1)}$ where $\left< \delta s^2 \right
>_\mu=\frac{1}{q_\mu}\sum_{i\in C_\mu} (\delta s_i)^2$ is the variance of
strength within $C_\mu$. In the first equation, we have used the
fact that $ ( \sum_{i \in C_\mu} \delta s_i )^2 = 2\sum_{i < j \in
C_\mu } \delta s_i \delta s_j + \sum_{i \in C_\mu } (\delta s_i) ^2=
0.$ The second equation holds simply because $\sum_{i \in C_\mu, j
\in C_\nu }\delta s_i \delta s_j = (\sum_{i \in C_\mu} \delta s_i)(
\sum_{j \in C_\nu} \delta s_j)=0.$ Substituting Eq.(\ref{eqCGA_ANA})
into Eq.(\ref{eqCGH}), we can get
\begin{subequations}\label{eqCGH_ANA}
\begin{eqnarray}
 \bar H_1  &=& - \frac{1}{{S N}}\sum\limits_\mu { {S_\mu^2 \left( {1 - \Omega_\mu } \right)\sum\limits_\alpha
          {\frac{{\eta _{\mu,\alpha} (\eta _{\mu,\alpha} - 1)}}{2}} } } \\
 \bar H_2  &=& \frac{1}{{S N}}\sum\limits_{\mu,\nu( > \mu)} {S_\mu S_\nu } \sum\limits_\alpha  {\eta _{\mu,\alpha} \eta
_{\nu,\alpha} } ,
  \label{eq12}
 \end{eqnarray}
\end{subequations}

To compare the CG-Hamiltonian with the microscopic one, we now group
the micro-nodes with same state $\alpha$ inside $C_\mu$ as $C_{\mu,\alpha}$.
Clearly, the size of $C_{\mu,\alpha}$ is $\eta_{\mu,\alpha}$. As in
Eq.(\ref{eqCGH}), we can also split the micro-Hamiltonian $H$ into
two parts ,
\begin{equation}
 H = H_1 + H_2
\end{equation}
where $H_1$ and $H_2$ denote energy contributions from intra and
inter the CG-nodes respectively. With ANA, and noting the fact only
nodes with same states have interactions at the micro-level, one has
\begin{subequations}\label{eqMicH}
\begin{eqnarray}
 H_1 &=& - \sum\limits_\mu \sum\limits_\alpha  {\sum\limits_{i < j \in C_{\mu,\alpha} }} \frac{s_i s_j}{S N} \frac{\eta
_{\mu,\alpha} (\eta _{\mu,\alpha} - 1)}{2}   \\
 H_2 &=& - \sum\limits_{\mu,\nu( > \mu)} \sum\limits_\alpha  {\sum\limits_{ i \in C_{\mu,\alpha},j \in C_{\nu,\alpha} } }
\frac{s_i s_j}{S
 N} \eta _{\mu,\alpha} \eta _{\nu,\alpha}
 \end{eqnarray}
\end{subequations}

Following similar steps to obtain Eq.(\ref{eqCGH_ANA}), we may also
write $s_i=S_{\mu,\alpha}+\delta s_i$ (here node $i$ belongs to the
group $C_{\mu,\alpha}$) and Eqs.(\ref{eqMicH}) change to
\begin{subequations}\label{eqMicH_ANA}
\begin{eqnarray}
 H_1 &=& - \sum\limits_\mu {\sum\limits_\alpha  {\sum\limits_{i < j \in C_{\mu,\alpha} } {\frac{{(S_{\mu,\alpha}
          + \delta s_i )(S_{\mu,\alpha}  + \delta s_j )}}{{S N}} } } } \times \nonumber \\ & & \frac{{\eta _{\mu,\alpha} (\eta _{\mu,\alpha} -
          1)}}{2} \nonumber \\
         &=&  - \frac{1}{{S N}}\sum\limits_\mu {\sum\limits_\alpha  { {S_{\mu,\alpha}^2
         (1 - \Omega _{\mu,\alpha} )\frac{{\eta _{\mu,\alpha} (\eta _{\mu,\alpha} - 1)}}{2}} } }  \\
 H_2 &=&  - \sum\limits_{\mu,\nu( > \mu)} {\sum\limits_\alpha  {\sum\limits_{i \in C_{\mu,\alpha} ,j \in C_{\nu,\alpha} }
           {\frac{{(S_{\mu,\alpha}  + \delta s_i )(S_{\nu,\alpha}  + \delta s_j )}}{{S N}} } } }\times \nonumber \\ & & \eta _{\mu,\alpha} \eta
_{\nu,\alpha} \nonumber \\
         &=&  - \frac{1}{{S N}}\sum\limits_{\mu,\nu( > \mu)} { \sum\limits_\alpha
         {\eta _{\mu,\alpha} \eta _{\nu,\alpha} S_{\mu,\alpha} S_{\nu,\alpha}} }
 \end{eqnarray}
\end{subequations}
Here $\Omega_{\mu,\alpha} = \frac{{ \left< \delta s^2 \right
>_{\mu,\alpha} }}{S_{\mu,\alpha}^2 (\eta _{\mu,\alpha} - 1)}$ where $\left< \delta s^2 \right
>_{\mu,\alpha}=\frac{1}{\eta _{\mu,\alpha} }\sum_{i\in C_{\mu,\alpha}} (\delta s_i)^2$ is the variance of
strength within the group of nodes $C_{\mu,\alpha}.$

Comparing Eq.(\ref{eqCGH_ANA}) with Eq.(\ref{eqMicH_ANA}), the
discrepancy between the CG-Hamiltonian and the micro-Hamiltonian, is
given by
\begin{subequations}\label{eqCG_Mic}
\begin{eqnarray}
\bar H_1  - H_1 &=&  - \frac{1}{{S N}}\sum\limits_\mu
{\sum\limits_\alpha  { \frac{{\eta _{\mu,\alpha} (\eta _{\mu,\alpha}
- 1)}}{2} } } \times \nonumber \\ & & {[ {S_\mu^2 ( {1 - \Omega _\mu
} ) - S_{\mu,\alpha }^2 (1 - \Omega _{\mu,\alpha } )} ]}
 \\
\bar H_2 - H_2 & = &- \sum\limits_{\mu,\nu} {\sum\limits_\alpha
{\frac{\eta _{\mu,\alpha} \eta _{\nu,\alpha} ( S_\mu S_\nu -
S_{\mu,\alpha }S_{\nu,\alpha }) }{{S N}} }}
\end{eqnarray}
\end{subequations}

Obviously, for the exact \sCG algorithm where all the nodes inside a
given CG-node have same strengths, $\Omega_\mu =
\Omega_{\mu,\alpha}=0 \ \forall (\mu,\alpha) $ and
$S_\mu=S_{\mu,\alpha}$, hence $\bar H_1=H_1$ and $\bar H_2=H_2$. In
this case, all those microscopic configurations contributing to a
CG-configuration $\vec {\bm \eta}$ have exactly the same Hamiltonian
$H$, which also equals to the CG-Hamiltonian $\bar H$. Since the
constrained summation $\sum\nolimits '$ contains exactly $g(\vec
{\bm \eta})$ items, the numerators on both sides of Eq.(\ref{eqCSC})
are exactly equal, i.e., $g(\vec {\bm \eta}) e^{-\bar H/k_B
T}=\sum\nolimits 'e^{- H/k_B T} $. Since we can also write the
microscopic partition function as $Z=\sum_{\vec {\bm \eta}} (
\sum\nolimits ' e^{-H/k_BT})$, it is readily to show that the two
partition functions equal, $\bar Z = Z$. Therefore, the CSC,
Eq.(\ref{eqCSC}), exactly holds.


However, we should note that for a weighted network, the exact \sCG method is not practical, since the strength of a given node is generally not an integer. Therefore, usually one can only merge nodes with close strength together.
Let us analyze Eq.(\ref{eqCG_Mic})
again. The factor $\Omega_\mu$ scales as $\frac{ { \left< \delta s^2
\right >_\mu }} {S_\mu^2 q_\mu}$, hence if we merge many nodes with
similar strengths together, $\Omega_\mu \ll 1$ is expected to be
true. One may also expect that $\Omega_{\mu,\alpha}\ll 1$ for the
same reason. Therefore, the discrepancy between $\bar H$ and $H$
mainly depends on the difference between $S_\mu$ and
$S_{\mu,\alpha}$. Here, we note that the nodes with $\alpha$-state
flip with time during the simulation. In the equilibrium state, one
expects that $C_{\mu,\alpha}$ may scan throughout $C_\mu$ for many
times, such that $S_{\mu,\alpha}$ averaged over time is close to
$S_\mu$. Hence $(\bar H-H)/H$ averaged over long time could be
small. Note that if we merge nodes randomly, $\Omega_\mu \ll 1$ and
$\Omega_{\mu,\alpha} \ll 1$ will be violated and the above reasoning
should fail. We thus conclude that the pratical \sCG approach, by merging nodes with similar strength together, can satisfy the CSC approximately.

\section{Numerical Results}  \label{sec3}

To show the validity of our \sCG approach, we perform extensive
simulations on weighted SF networks. SF networks are much
heterogeneous and serve as better candidates to test our method than
other homogeneous networks, such as small-world or random networks
(other types of complex networks have also been investigated, the
qualitative results are the same and not shown here). We first
generate a regular (unweighted) SF network by using the
Barab\'{a}si--Albert (BA) model \cite{SCI99000509} with power-law
degree distribution $P(k)\sim k^{-3}$. To convert this unweighted SF
network into a weighted one, we use the algorithm as proposed in
Ref. \cite{PRE04026109}: The weight of a link between node $i$ and
$j$ ($1\le i,j\le N$) is given by  $ w_{ij} =(
\frac{i}{N}+\frac{j}{N} )^\theta / 2$, where $\theta$ is a tunable
parameter. Note that $\theta = 0$ corresponds to an unweighed
network.

The MC simulation at the microscopic level follows standard Metropolis dynamics: At each step, a micro-node is randomly selected and its state is randomly updated
with an acceptance probability $\min(1, e^{-\Delta H/k_B T})$, where
$\Delta H$ is the associated change of the micro-Hamiltonian, $k_B$ is the Boltzmann constant, $T$ is the temperature. In the present work, we set
$k_B=1$. Similarly, during each CG-MC step, a CG-node $C_\mu$ is randomly
chosen with probability proportional to its size $q_\mu$. The
probability for the process that an $\alpha$-node changes to a
$\beta$-node, with correspondingly $\eta_{\mu,\alpha}\rightarrow
\eta_{\mu,\alpha}-1$ and $\eta_{\mu,\beta} \rightarrow
\eta_{\mu,\beta}+1$, is given by $\eta_{\mu,\alpha} \min(1, e^{ -
\Delta \bar H/k_B T} )$, where $\Delta \bar H $ is the change of
CG-Hamiltonian during this process. Since $N^c$ can be much smaller
than $N$, the CG-MC is expected to be much faster and memory-saving
than the micro-level MC simulation.

The collective state of the system is described by the total
magnetic moment $ M = \frac{1}{{2N}}\sum\limits_{\mu,\alpha }
{\left| {M_{\mu,\alpha } } \right|}$, where $ M_{\mu,\alpha }  =
\frac{{p\eta _{\mu,\alpha } - 1}} {{p - 1}}{\kern 8pt} ( \mu= 1,
\cdots ,N^c)$ denotes the $\alpha$-component of the magnetic moment
within $C_\mu$. With increasing temperature $T$, the Potts model
undergoes a phase transition at some critial temperature $T_c$ from
an ordered state, where $M \sim O(1) $ is strictly nonzero, to a
disordered state with $M \simeq 0$. We use the similar \sCG approach
to construct the CG-network with different $N^c$ and compare the
results obtained from CG-MC simulations with those of micro-MC
simulations.

 To begin, we show the results in Fig.1 for $\theta = 0$, where the network are essentially unweighted and the \sCG approach is identical to the \dCG. Fig.1(a) and 1(b) show the moment $M$ and susceptibility $\chi=\beta N(\langle M^2 \rangle - \langle M\rangle ^2 )$ as functions of $T$, respectively. The susceptibility is related to the variance of the total
magnetization according to the fluctuation-dissipation theorem.
Apparently, our results (empty squares and solid circles) are in
excellent agreements with the micro-level counterparts (solid
lines). As comparisons, we have also shown the results obtained by a
random-merging (RM) CG-model (dotted lines) and the heterogeneous
mean field theories (HMFT)\cite{EPJ04000177} (empty triangles).
Here, the RM model means that one simply merge $N/N^c$ randomly
selected nodes to form a CG-node. Evidently this random scheme fails
to reproduce the microscopic behaviors at all. The results of the
HMFT are obtained by numerically solving the self-consistent
equations of order parameter\cite{EPJ04000177}. We find that the
HMFT can predict the curve of $M \sim T$ quite well, however, it
fails to predict the curve of $\chi \sim T$. Strikingly, even when
the original network is reduced to one with only $16$ CG-nodes, the
CG model still faithfully reproduces the phase transition curves and
fluctuation properties. Since $N^c$ is largely reduced compared to
$N$, a considerable speed-up of CPU time can be achieved which makes
it feasible to study system size effects.  Fig.1(c) plots $T_c$ as a
function of  $\ln N$, obtained by our CG method with $N^c=64$. $T_c$
is determined as the location of the peak in the $\chi \sim T $
curve, see Fig.(1b). The dependence is linear with a slope $\simeq
1.68$, which agrees rather well with a theoretical prediction
${T_c}/\ln N=\frac{S}{4p}\simeq1.67$ \cite{EPJ04000177}, where $S$
is the average node strength in the network.

\begin{figure}[h]
\centerline{\includegraphics*[width=1.0\columnwidth]{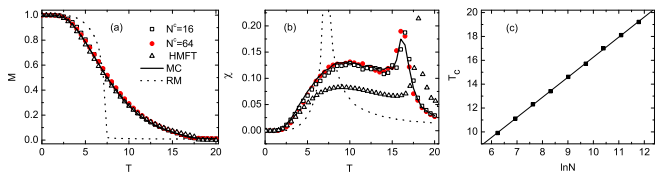}}
\caption{(color online). (a)-(b) $M$ and $\chi$ as functions of $T$
(in unit of $J/k_B$) for the Potts model on unweighted SF networks
($\theta =0$), obtained from brute-force MC simulation (solid line),
HMFT(triangle), random-merging CG(dotted line), and the \sCG (square
and circle). $N=16384$, $p=3$. (c) Dependence of $T_c$ on the
network size $N$ obtained by the \sCG approach with fixed $N^c=64$.
All the networks have fixed mean degree $D=20$. The error bars (not
shown) are smaller than the symbol sizes. \label{fig1}}
\end{figure}

For $\theta \ne 0$, the networks are weighted. Here we take $\theta
=2.4$ as an example to ensure the heterogeneity of the link weights.
Figure(2a) and (2b) show $M$ and $ \chi $ as functions of $T$
respectively. As in Fig.1(b), the peak in $\chi$ locates the
critical point $T_c$. Clearly, the \sCG results (solid circle) are
still in excellent agreements with the MC results (solid lines),
however, the \dCG (solid squares) \cite{PRE10011107} and RM-CG
(dotted lines) both fails. For such weighted networks, the dynamic
equations of HMFT is not available either. Thus, for such weighted
networks, our \sCG approach is the only promising CG approach so
far. In Fig.2(c), we have also shown the dependence of $T_c$ on the
network size. Apparently, there is also a linear dependence between
$T_c$ and $\ln N$ with the slope being about $1.288$. As mentioned
in the last paragraph, this slope depends on the average strength
$S$. For a weighted network, one may estimate $S$ by $\langle w_{ij}
\rangle D$, where $\langle w_{ij} \rangle  \simeq \int_0^2 x^\theta
/4dx = \frac{1}{{4(\theta + 1)}}2^{\theta  + 1} $. Substituting $ D
= 20$, $\theta = 2.4$ and $p = 3$ to these formula, we obtain
${T_c}/\ln N=\frac{S}{4p}\simeq1.293$, which is consistent with the
simulation value.

\begin{figure}[h]
\centerline{\includegraphics*[width=1.0 \columnwidth]{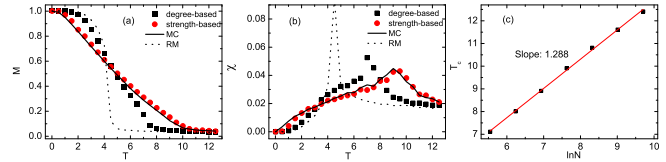}}
\caption{(color online). (a)-(b) $M$ and $\chi$ as functions of $T$
for the ferromagnetic Potts model on a weighted scale-free network
with mean degree $D=20$. $N=1024$, $p=3$, $\theta =2.4$ and
$N^c=16$. (c) Dependence of $T_c$ on the network size $N$. The error
bars are omitted for clarify since they are smaller than the symbol
sizes.}
\end{figure}

In real-world networks, correlation is an ubiquitous feature. For
instance, social networks show that nodes with large degrees tend to
connect together, a property referred to as \lq\lq assortative
mixing\rq\rq\cite{PRL02208701}. In contrast, many technological and
biological networks show \lq\lq disassortative mixing\rq\rq, i.e.,
connections between high-degree and low-degree nodes are more
probable \cite{PRL01258701,SCI02000910}. Previous studies showed
that correlations may play important roles in network dynamics
\cite{PRL02208701,PRL01258701,SCI02000910,PRE08051105,PRE02047104}.
In the present work, we have used our \sCG method to study the phase
transition of Potts model on correlated networks, which can not be
studied by the HMFT which assumes no degree
correlation. To characterize the assortative property of the weighted
network, a strength correlation coefficient $r$, an extension of the
degree correlation \cite{PRL02208701}, can be defined as
\begin{equation}
r  = (\langle s_i s_j \rangle  - \langle s_i \rangle \langle s_j
\rangle )/(\langle s_i^2 \rangle  - \langle s_i \rangle ^2 ) .
 \label{eq13}
\end{equation}
Here $s_i$ and $s_j$ are the strengths of the two end-nodes of an
 edge. $r$ is zero for networks with no strength-correlation,
such as BA-SF networks, and positive or negative for assortative or
disassortative mixing networks, respectively.

\begin{figure}[h]
\centerline{\includegraphics*[width=0.9\columnwidth]{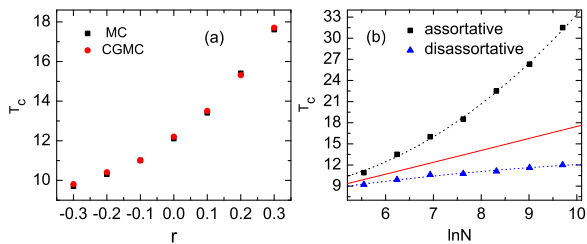}}
\caption{(color online). Phase transition behaviors of the Potts
model on unweighted correlated networks. (a) $T_c$ plotted as a
function of the network correlation coefficient $r$, obtained via
CG-MC and MC simulations. $N=1024$ and $N^c=64$. (b) Dependence of
$T_c$ on the network size $N$. All the networks have fixed mean
degree $D=20$. \label{fig3}}
\end{figure}

Figure 3(a) shows $T_c$ as a function of $r$, obtained from our \sCG
approach and micro-MC simulations for $\theta = 0$. Again, the fits
between CG-MC and MC are good. Figure 3(b) shows the effects of
correlated network size on $T_c$. Interestingly, we find that the
linear dependence between $T_c$ and $\ln N$ is lost for correlated
networks. For assortative(disassortative) networks $T_c$ grows
monotonically much faster(slower) than $\ln N$, respectively. In
other words, the ordered state in an assortative(disassortative)
network is harder(easier) to be destroyed with increasing
temperature than in an un-correlated network. This is understandable
since a \lq hub\rq-node in the network is more difficult to change
its state than a \lq leaf\rq-node due to larger energy barrier. In
an assortative network, hub-nodes are connected together, such that
they tend to freeze into a local ordered state which is stable to
thermal fluctuations. For a disassortative network, a hub-node is
usually connected to many leaf-nodes. Since leaf-nodes can change
state easily, the \lq alone\rq ${\kern 1pt}$ hub-node is more likely
to change state with the help of their \lq boiling\rq ${\kern 1pt}$
neighbors. Therefore, assortative correlations tend to increase
$T_c$ as observed here.

\begin{figure}[h]
\centerline{\includegraphics*[width=0.9\columnwidth]{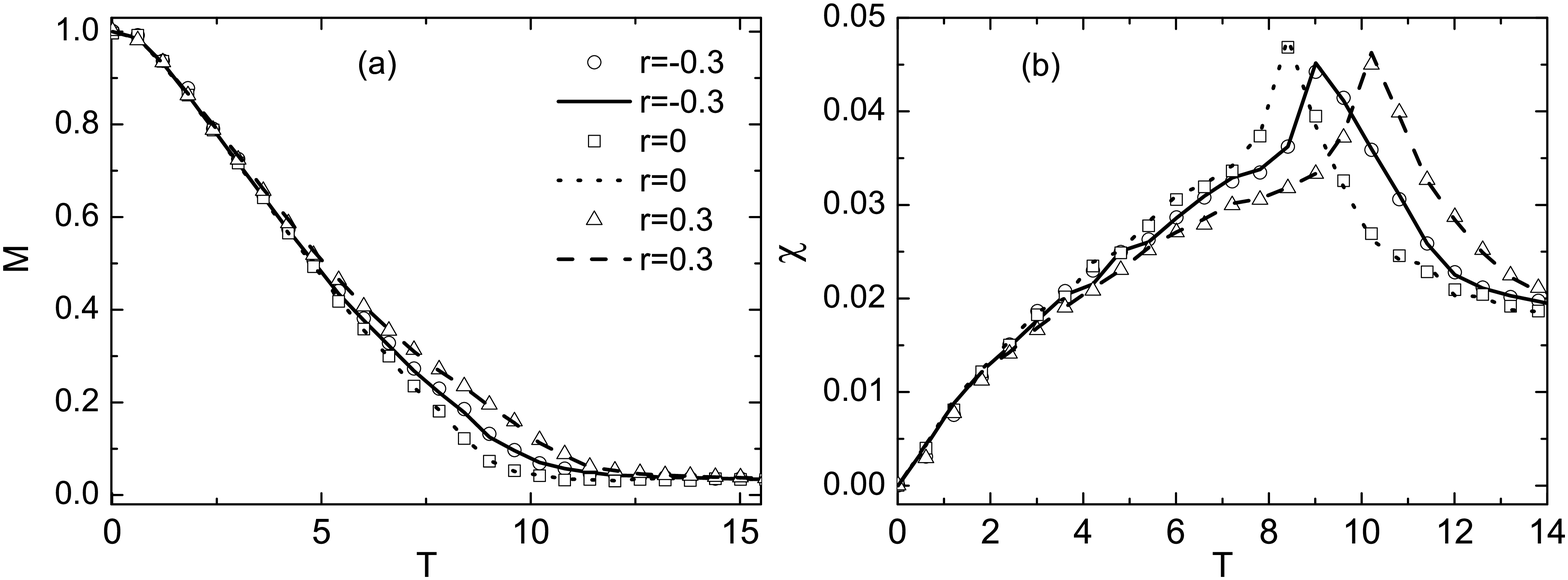}}
\caption{ Magnetization $M$ and susceptibility $\chi$ as functions
of temperature $T$ for the Potts model on weighted correlated
networks. Symbols and lines correspond to the CG-MC and micro-MC
simulation results, respectively. Other parameters are same as in
Fig.2. \label{fig4}}
\end{figure}

In Fig.(4), the magnetization $M$ and susceptibility $\chi$ of the
ferromagnetic Potts model on weighted networks are plotted as
functions of temperature $T$ at different correlation coefficient
$r$, obtained from our \sCG approach and micro-MC simulations.
Again, the agreements between CG-MC and MC are excellent, further
demonstrating the validity of our method.

\section{Conclusions} \label{sec4}

In summary, we have developed a stength-based \sCG approach for
coarse-graining study of the phase transition of the Potts model on
weighted networks. We have utilized a mean-field scheme to generate
the connectivity of the CG-network and derived the CG-Hamiltonian.
To address the problem how to guarantee the validity of the
CG-model, we have proposed the so-called CSC, which requires that
the probability to find a given CG-configuration in the equilibrium
state, calculated from the CG-model, should be the same as that
calculated from the original microscopic model.  We show, by
performing error analysis, that our \sCG approach, by merging nodes
with close strengths together, holds the CSC approximately with ANA.
Detailed numerical simulations demonstrate clearly that our \sCG
approach can reproduce the microscopic MC simulation results very
well, not only for the onset of phase transition, but also for the
fluctuations and system size effects.

\begin{acknowledgments}
This work was supported by the National Science Foundation of China
under Grants No.20933006 and No.20873130.

\end{acknowledgments}

%

\end{document}